\documentclass{article}

\PassOptionsToPackage{numbers, compress}{natbib}

    \usepackage[preprint]{neurips_2024}

\usepackage[utf8]{inputenc} 
\usepackage[T1]{fontenc}    
\usepackage{hyperref}       
\usepackage{url}            
\usepackage{booktabs}       
\usepackage{amsfonts}       
\usepackage{nicefrac}       
\usepackage{microtype}      
\usepackage{xcolor}         
\usepackage{amsmath,amsfonts,amssymb,amsthm}        
\usepackage{graphicx}
\usepackage{tabularx}
\usepackage{multirow} 
\usepackage{amsmath}
\usepackage{bm}
\def\softmax {\mathrm{softmax}}

\def\pr#1{\left( #1\right)}

\def\ang#1{\left\langle #1\right\rangle}

\newcommand{\iu}{\mathrm{i}\mkern1mu}

\title{Dynamical Mean-Field Theory of Self-Attention Neural Networks}

\author{%
  Ángel Poc-López
    \\
  BCAM -- Basque Center for Applied Mathematics.\\
  Bilbao, Spain 48009 \\
  Department of Computer Science and Systems Engineering (DIIS), University of Zaragoza. \\
  Zaragoza, Spain 50018 \\
  \texttt{apoc@bcamath.org} \\
  \And
  Miguel Aguilera \\
  BCAM -- Basque Center for Applied Mathematics.\\
  IKERBASQUE, Basque Foundation for Science.\\
  Bilbao, Spain 48009 \\
  \texttt{maguilera@bcamath.org} \\
}

\begin{document}

\maketitle

\begin{abstract}

Transformer-based models have demonstrated exceptional performance across diverse domains, becoming the state-of-the-art solution for addressing sequential machine learning problems. Even though we have a general understanding  of the fundamental components in the transformer architecture, little is known about how they operate or what are their expected dynamics. Recently, there has been an increasing interest in exploring the relationship between attention mechanisms and Hopfield networks, promising to shed light on the statistical physics of transformer networks. However, to date, the dynamical regimes of transformer-like models have not been studied in depth. In this paper, we address this gap by using methods for the study of asymmetric Hopfield networks in nonequilibrium regimes --namely path integral methods over generating functionals, yielding dynamics governed by concurrent mean-field variables. Assuming 1-bit tokens and weights, we derive analytical approximations for the behavior of large self-attention neural networks coupled to a softmax output, which become exact in the large limit size. Our findings reveal nontrivial dynamical phenomena, including nonequilibrium phase transitions associated with chaotic bifurcations, even for very simple configurations with a few encoded features and a very short context window. Finally, we discuss the potential of our analytic approach to improve our understanding of the inner workings of transformer models, potentially reducing computational training costs and enhancing model interpretability. \footnote{\tiny{Code available at: \url{https://github.com/muxitox/Dynamical-MF-Self-Attention}}}
\end{abstract}

\section{Introduction}


Transformer-based approaches have demonstrated enormous capabilities in terms of performance and quality in different domains and modalities \cite{vision_transformers, udandarao2024zeroshot}. Nevertheless, their functional dynamics remain largely mysterious. Transformers are typically crafted through a process of trial and error, and their enormous architectures need to be fed with exponential quantities of data in order to reach zero-shot performance \cite{udandarao2024zeroshot}. Moreover, the resulting behavior is challenging to interpret and forecast, presenting substantial challenges concerning AI alignment  \cite{russell2019human}.
Given these challenges, we propose that methods from nonequilibrium statistical mechanics offer a promising avenue for elucidating the behavior of large transformer models (e.g., the dynamics of emerging order parameters and their phase transitions), as well as in alleviating computations that currently depend on conventional brute-force statistical methods.

Much of our understanding of the statistical physics of neural networks dynamics comes from Hopfield networks \cite{amit1989modeling}, a well-established  model exemplifying how simple networks can exhibit complex behaviors reminiscent of memory storage and retrieval \cite{hopfield1982neural}.  
As a conceptual framework, Hopfield networks have profoundly impacted several disciplines, spanning statistical physics, neuroscience, and machine learning \cite{peretto1984collective,hillar2018robust,furst2022cloob}.
Recently, there has been an upsurge in the machine learning community's interest with this model. This renewed enthusiasm stems from the striking parallels observed between attention mechanisms in transformers and Hopfield networks \cite{ramsauer2020hopfield}. 
More generally, attention can be mapped to classes of network models in statistical mechanics, like Ising and Potts models \cite{rende2024mapping}, offering new avenues for exploring theoretical principles of the operation and design of transformer innovative architectures \cite{krotov2023new}.
The link with attention networks has generally been established by drawing parallels between the \emph{softmax} function in attention and a specific version of Hopfield networks referred to as \emph{modern Hopfield networks} (i.e. continuous-state instead of binary) with a \emph{LogSumExp} energy function (integral of the \emph{softmax} function) \cite{ramsauer2020hopfield,krotov2023new}. Nonetheless, the equivalence can be obtained --more generally-- by simply describing a Hopfield network encoding a conditional probability distribution \cite{rende2024mapping,singh2023attention}.

Here, we claim that a more general equivalence can be obtained by modelling an attention layer as a \emph{nonequilibrium} (i.e., with asymmetric couplings) Hopfield network \cite{treves1988metastable}, driven by simple quadratic couplings. More importantly, this description allows us to use tools from nonequilibrium statistical physics for solving the network's sequential dynamics. As an example of this approach, we introduce an array of mean-field variables solving the recurrent dynamics of a decoder attention layer connected to a linear softmax output.
Our calculations, which are exact in the large limit size (i.e., the number of neurons approaches infinity), reveal the existence of chaotic phase transitions even in very simple configurations of the model, with only a few encoded features and a context window of a few time-steps.
To our knowledge, this is the first systematic description of dynamical regimes (i.e., the evolution of the system's sufficient statistics over time) of a simplified transformer-based model and its phase transitions. 

We proceed as follows. First, we establish the equivalence between attention networks and asymmetric Hopfield networks. As well as a similar equivalence between an asymmetric Hopfield network and a simplified \emph{softmax} output block.
Then, we will derive a dynamical mean-field theory describing the statistics of path probabilities from a set of low-dimensional mean-field variables, representing the overlap of the system's state and different features.
Using this analytical solution, we explore the resulting dynamics in some simple examples, exploring the phase diagram of the system. Finally we discuss the relevance of our framework for the understanding of attention networks, as well as the possibilities of this research direction regarding model training and interpretability.

\section{Hopfield networks and transformers}

A Hopfield network \cite{hopfield1982neural} describes a system in which the probability $p(\bm x)$ of a set spins $x_i \in \{-1,+1\}$ for $i\in\{1,..,N\}$ is defined though an energy function which takes into account the $M$ memories stored in the system $\pmb{\xi}_a=\{ \xi_{a,0} , .., \xi_{i,a}, ..,\xi_{a,N}\}$ for $a\in\{1,..,M\}$ where $\xi_{i,a} \in \{-1,+1\}$. The probability of a state is defined as
\begin{align}
    p(\bm x) &= Z^{-1} \exp\Big(\frac{\beta}{N} \sum_a \sum_{i<j}  x_i \xi_{i,a} \xi_{j,a} x_j\Big).
    \label{eq:hop-net}
\end{align}
Here, $ Z = \sum_{\bm{x}} \exp(\beta\sum_a \sum_{ij}  x_i \xi_{i,a} \xi_{j,a} x_j )$ is a partition function and couplings are usually symmetric (coupling between neuron $i$ and $j$ has a value of $\sum_a \xi_{i,a} \xi_{j,a} $), describing an energy landscape $E$, in which minima we can recover the memories of the system. The parameter $\beta$ is a constant defining the inverse temperature. Finally, the normalization $\frac{1}{N}$ term ensures that the energy of the system is extensive.

Similarly to restricted Boltzmann machines \cite{hinton2012practical}, we can describe a bipartite Hopfield network by defining two sets of variables $\bm x = \{\bm k, \bm q\}$ and a series of patterns encoded by matrices $\bm W^k, \bm W^q$
\begin{align}
    p(\bm k,\bm q) &= Z^{-1} \exp\Big(\frac{\beta}{N} \sum_a \sum_{ij} k_i W^k_{i,a} W^q_{j,a} q_j\Big). \label{eq:bipartite_Hopfield}
\end{align}
Note that now connections between $\bm k,\bm q$ are asymmetric (i.e., potentially, $\bm W^k_{a} \neq \bm W^q_{a} $).
For a given $\bm q$, the conditional distribution of the system can be obtained as
\begin{align}
    p(\bm k|\bm q) =&\frac{\exp\Big(\dfrac{\beta}{N} \sum_a \sum_{ij}  k_i W^k_{i,a} W^q_{j,a} q_j\Big)}{\sum_{\bm k'} \exp\Big(\dfrac{\beta}{N} \sum_a \sum_{ij}  k'_i W^k_{i,a} W^q_{j,a} q_j\Big)}
     \label{eq:bipartite_Hopfield_conditional}
\end{align}
When the output $\bm k$ is fed again into the input $\bm q$, symmetric random couplings give rise to spin glass behaviour \cite{brunetti1992asymmetric}, whereas asymmetric random couplings give rise to a nonequilibrium steady states displaying order-disorder phase transitions and chaotic dynamics  \cite{eissfeller1994mean,aguilera2023nonequilibrium}.

\begin{figure}
    \centering
\begin{tabular}{c c c} \includegraphics[width=0.2\linewidth]{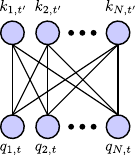} & \hspace{1cm} &\includegraphics[width=0.2\linewidth]{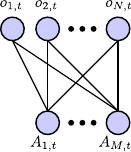} \\
        a) &  &b) 

\end{tabular}
    \caption{a) Bipartite Hopfield network, equivalent to the attention mechanism when summing over $\tau \in \{1,2,\dots,t\}$. b) Description of a \emph{softmax} output with a linear encoding over attention values.}
    \label{fig:bipartite_Hopfield}
\end{figure}

If we look closely at the self-attention function defined in \cite{luong2015effective,vaswani2017attention} for a unique head,  $\bm{A}_t = \sum_{\tau} \softmax(\frac{1}{U}(\bm q_t \bm W^q)^\intercal \bm W^k \bm k_{t-\tau})_\tau \bm W^v \bm v_{t-\tau}$, we can observe that the query key interaction matrices have a similar form as the Hopfield network defined above. Introducing a scaling parameter $\gamma$  (similar to the inverse temperature $\beta$ above, but we reserve this term for the output later), we describe
\begin{align}
    \notag\\ A^a_t =& \sum_{\tau} p_A(\bm k_\tau|\bm q_t) \sum_i  W_{i,a}^v  v_{i,t-\tau},
    \label{eq:attention}
    \\ p_A(\bm k_\tau|\bm q_t) = &\, \softmax_\tau\Big(\dfrac{\gamma}{U} \pr{\bm W^q\bm q_t }^\intercal \bm W^k \bm k_{t-\tau}\Big)  
    \nonumber\\=& \frac{\exp\Big(\dfrac{\gamma}{U} \sum_a \sum_{ij}  k_{j,t-\tau} W_{j,a}^k W_{i,a}^q q_{i,t}\Big) }{\sum_{\tau'} \exp \Big(\dfrac{\gamma}{U} \sum_a \sum_{ij}  k_{j,t-\tau} W_{j,a}^k W_{i,a}^q q_{i,t}\Big) }.
    \label{eq:softmax-attention}
\end{align}
where the sums are over $\tau \in \{0,1,\dots, L\}$, being $L$ the length of the context window, and $a\in\{1,..,M\}$ defines the encoded features as in Eq.~\eqref{eq:hop-net}. Importantly, in self-attention, each key, query and value tokens are equal to inputs different time steps $\bm x_t$ so $\bm k_t = \bm q_t = \bm v_t =  \bm x_t \;\forall t$, as described in Eq.~\eqref{eq:attention-x} (in contrast, in cross attention query tokens can take the value of a different input).
Notice that we have opted for expressing the softmax with a partition function, obtaining an equivalent form to the conditional distribution of an asymmetric Hopfield network in Eq.~\eqref{eq:bipartite_Hopfield_conditional}. We note that the equivalence is valid even in the case where vectors $\bm x$ are not limited to binary values.
Importantly, attention presents two important differences respect to the bipartite Hopfield network. First, the summation of the partition function is not over all queries $\bm q$, but only those that appeared at previous steps $t-\tau'$, as we observe in Eq.~\eqref{eq:softmax-attention}. Second, the dot product is normalized so that the softmax function  is always in regions with relatively large gradients, assuming that the variance of $\sum_i q_{i,t} W_{i,a}^q,\sum_i k_{i,t} W_{i,a}^k$ terms over $a$ is around order 1 \cite{vaswani2017attention}. Generally, unit variance is satisfied through dedicated normalization layers and weights initialized so that variance is preserved. For a large number of features, the softmax employs a normalization term  $U = \sqrt{M}$. In our case, for simplicity we prescind from normalization layers and we include weight normalization into our normalizing constant to get $U = N^2 \sqrt{M}$.

\subsection{Simplified attention-output layer}


Generally, the output in a transformer network is defined as a softmax. Our goal is mostly to devise the types of sequential behavior a self-attention layer is able to produce. As a consequence, we are opting for representing a highly simplified transformer network as the combination of an attention layer and a softmax output. Similar single-layer attention networks with a nonlinear output can completely
memorize finite samples and are universal approximators for continuous functions \cite{kajitsuka2023transformers}.
To focus on the dynamics of attention, we ignore the addition, normalization and feed-forward network blocks, and apply the softmax directly to logits defined as a linear combination of the attention values as $p(\bm o_{t}|\bm A_{t}) = \softmax_{\bm o_t}(\bm{A}_{t}^\intercal \bm W^o_{\bm o_t})$. To further simplify the model, we assume that the output logits are obtained as a linear transformation $\bm W^o_{\bm o} = \bm W^o \bm o $. Then
\begin{align}
    p(\bm o_{t}|\bm A_{t}) =& 
    \softmax_{\bm o_t} (\frac{1}{N} \bm{A}_{t}^\intercal \bm W^o \bm o_t) 
    \nonumber\\=& \frac{\exp\Big( \frac{\beta}{N} \sum_a \sum_i o_{i,t} W^o_{i,a} A_t^a \Big)}{\sum_{\bm o'}\exp\Big( \frac{\beta}{N} \sum_a \sum_i o'_{i,t} W^o_{i,a} A_t^a\Big)},
    \label{eq:softmax-output}
\end{align}
where $\beta=T^{-1}$ is the inverse temperature, and $\frac{1}{N}$ normalisation ensures unit variance of logits.

\begin{figure}
    \centering
    \includegraphics[width=0.4\linewidth]{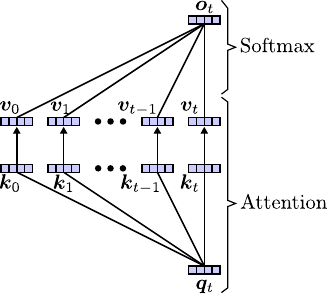}
    \caption{Description of an attention layer connected to a \emph{softmax} output. Each link between queries $k_t$ and keys $k_t$, as well as between attention values $v_t$ and outputs $o_t$, follow represent a bipartite Hopfield network as in Fig.~\ref{fig:bipartite_Hopfield}. Attention values $v_t$ take the same value as $k_t$.}
    \label{fig:attention_diagram}
\end{figure}



To generate the transformer dynamics, we define sequences of states $\bm x_{0:T} = \{\bm x_0, \bm x_1, \ldots, \bm x_T\}$, with $\bm x_t = \{x_{1,t},\dots,x_{N,t}\}$.
We then define the dynamics of the model generating attention values from Eqs.~(\eqref{eq:attention}, \eqref{eq:softmax-attention}) with $\bm v_\tau = \bm k_\tau$ and later generating the next token $\bm x_{t+1} = \bm o_t$ autoregressively using Eq.~\eqref{eq:softmax-output} as $p(\bm x_{t+1}|\bm A_{t})$, yielding
\begin{align}
    p(\bm x_{0:T}) =& \prod_{t=0}^{T-1} p(\bm x_{t+1}|\bm A_{t}), \label{eq:path-probability}
    \\  A_{t}^a =& \sum_{\tau} p_A(\bm x_\tau|\bm x_t)  \sum_i W^v_{i,a} x_{i,t-\tau}.
    \label{eq:attention-x}
\end{align}

\section{Dynamical mean-field theory of attention layers with 1-bit token encodings}

In this section, we study the recurrent dynamics of our simplified transformer using standard methods for examining the nonequilibrium statistical physics of recurrent networks \cite{coolen2001statistical}. 
For simplicity, and maintaining a description in terms of Hopfield models, we assume a binary encoding of weights and tokens. Binary weight transformers are quickly becoming an attractive alternative to implement significantly more cost-effective models \cite{ma2024era,wang2023bitnet}. Binary tokens are less common, but have been explored to reduce computational costs \cite{liu2022bit,qin2022bibert} (although effective scaling seems to be a key factor to maintain performance \cite{zhang2024binarized}). In any case, our framework can be easily extended to discrete sets of non-binary tokens (e.g. substituting $\tanh$ functions in our results by softmax).

Since the interaction between patterns is not symmetric as in Eq.~\eqref{eq:hop-net}, we do not have an analytic expression of the free energy from which to compute the moments of the system. Instead, we define a generating functional, which behaves as a moment generating function, playing an equivalent role to the partition function in equilibrium statistical mechanics but applicable to nonequilibrium settings. The generating functional is defined as:
\begin{align}
    Z(\bm g) =& \sum_{\bm x_{0:T}}  p(\bm x_{0:T}) \exp( \sum_{a,t}\sum_\alpha g_{a,t}^\alpha \frac{1}{N} \sum_i W^{\alpha}_{i,a} x_{i,t} ) ,
    \label{eq:generative-functional}
\end{align}
where $\alpha \in \{q,k,v,o\}$ is the index of the features of queries, key, value and output variables, and the path probabilities are defined in Eq.~\eqref{eq:path-probability}. Note that
\begin{align}
    \left.\frac{\partial Z(\bm g)}{\partial g_{a,t}^\alpha}\right|_{\bm g = \bm 0} = \frac{1}{N}\sum_i W^{\alpha}_{i,a}\ang{x_{i,t}}
\end{align}
recovers the statistics of the features encoded in the transformer. For this reason, the generating functional behaves as a sort of \emph{dynamical partition function} describing processes in nonequilibrium statistical physics \cite{coolen2001statistical}.

We solve the system using path integral methods \cite{coolen2001statistical} introducing mean-field variables 
\begin{align}
    m_{a,t}^\alpha = \frac{1}{N}\sum_i W^{\alpha}_{i,a} \ang{x_{i,t}} ,
\end{align}
calculating  Eq.~\eqref{eq:generative-functional} using a  steepest descent in the large limit size (Appendix~\ref{sec:appendix-functional}). In the case of 1-bit token encodings, this results in a generating functional
\begin{align}
    Z(\bm g) =&   \exp \Bigg( \sum^T_{t=0} \sum_i \log 2 \cosh \Big(\sum_a W^o_{i,a} \beta \hat{A}_{a,t-1} +  \frac{1}{N} \sum_{a,\alpha} W^\alpha_{i,a} g^\alpha_{a,t-1}\Big) 
    \notag \\&- \log 2 \cosh \Big(\beta  \sum_a W^o_{i,a} \hat{A}_{a,t}) \Big) \Bigg) 
\end{align}
described by the following equations for mean-field variables:
\begin{align}
        m^\alpha_{a,t} =& \frac{1}{N} \sum_i W^\alpha_{i,a}  
             \tanh (  \beta \sum_b  W^o_{i,b}  \hat{A}_{b,t-1}),
        \label{eq:mean-field}
   \\ \hat{A}^a_t =& \frac{ \sum_{\tau} m^v_{b,t-\tau} \exp (\gamma \sum_{a} m^q_{a,t} m^k_{a,t-\tau})}{\sum_{\tau'} \exp ( \gamma  \sum_{a} m^q_{a,t} m^k_{a,t-\tau'})}. 
    \label{eq:mean-field_att}
\end{align}
where $\hat{A}^a_t $ is the normalized attention value (i.e., divided by $N$) evaluated for the mean-field variables.
These equations take the familiar form of dynamical mean-field theory in asymmetric Hopfield networks, but they add an attention term in the form of a softmax over the mean-field values encoding overlaps with the model features. These equations are exact in the large limit size, though we should expect thermal fluctuations at smaller sizes.

We can observe that for all mean-fields $m^\alpha$ the solution takes the same form. In Eq.~\eqref{eq:mean-field} we see that the average behavior of spin $x_{i,t+1}$ for all patterns $b$ is computed within the $\tanh$ function to later be projected against each mean-field associated matrix $W^{\alpha}_a$, for the pattern $a$ to be computed. The information of all spins $i$ is gathered to obtain the total behavior of pattern $a$. We  may observe, however, that such network does not possess any information about token ordering. Transformer models alleviate this by adding a positional encoding in the form of an external signal \cite{vaswani2017attention}. To this end, we add $N_P$ units carrying the positional encoding to our tokens (i.e., the semantic embedding). In this embedding, we simply codify the information about the position of the token selected at time $t$ as an array of bits. We define positional token units as $p_{i,t}= (-1)^{\lceil t/i \rceil}$ (with $\lceil x \rceil$ being a ceil operator), being the $i$-th bit in a binary encoding of the time sequence value $t$
\begin{align}
        m^\alpha_{a,t+1} =& (1-\epsilon) \frac{1}{N} \sum_{i=1}^N
        W^\alpha_{i,a} 
        \tanh (  \beta \frac{1}{U} \sum_b  W^\alpha_{i,b}  \hat{A}_{b,t}) + \epsilon \frac{1}{N_P}\sum_{i=1}^{N_P} W^\alpha_{i,a} p_{i,t+1},
    \label{eq:mf_o_pe}
    \end{align}
where $\epsilon$ determines the relative weight of the positional encoding. The positional encoding in the equation above is just added to the tokens (or mean-fields) generated by the softmax output. Notice that the solution from Eq.~\eqref{eq:mean-field} is still correct but we are just recalculating its value adding an external signal $\bm p_t$.
As usual in transformer networks, the positional and semantic embeddings are projected through the same matrix $\bm W^\alpha$.

The mean-field equations in Eq.~\eqref{eq:mf_o_pe} cannot be directly calculated in the limit of infinitely large networks. Nevertheless, for a limited number of patterns encoded in $\bm W$ with values $\pm 1$, we can substitute the sums over $i$ and just use the correlation values between pairs of patterns $\bm W_a^\alpha, \bm W_b^{\alpha'}$ as shown in Eq.~\eqref{eq:mean_field_corrs} from Appendix \ref{sec:appendix-correlations}, resulting in:
\begin{align}
m^\alpha_{a,t} =& 
\frac{1}{2^M} \sum_{\boldsymbol{\sigma}}\Bigg(\Big( \sum_b \sigma_b \ang{W^o_{i,b} W^\alpha_{i,a}}_i + \sum_{b<c<d}  \sigma_b \sigma_c \sigma_d \ang{W^o_{i,b}W^o_{i,c}W^o_{i,d} W^\alpha_{i,a}}_i  + \cdots\Big)
\nonumber\\ &\cdot
\tanh \Big(\beta \sum_b \sigma_b \hat{A}_{b,t-1} \Big)\Bigg).\label{eq:mean_field_C}
\end{align}
where $\bm\sigma$ is an array of $M$ binary variables $\sigma_a = \pm 1$ and $\ang{\dots}_i$ indicates an average over the $i$ index. The dots include odd products of components of sigma times the average over $i$ of $W_{i,a}^\alpha$ multiplied by even products of components of $\bm W_{i}^o$.

\section{Results}


With the aim of observing the dynamical regimes of behavior of our simplified transformer network, we simulated different networks with random binary weights with also random correlation values (Appendix \ref{sec:appendix-corr-initialization}) and using a context window of $L=4$ tokens. We simulate the network for $1.2 \cdot 10^6$ steps, where each step takes as  input the last 4 tokens of the trajectory as defined by the context window, and generates the next token. To avoid transient trajectories, we discard the first $10^6$ steps. 
In the next sections, we will show the behavior of one of such network initializations.  Even one configuration of parameters allows to observe very different behaviors, and similar results can be reproduced for other combinations. The results obtained in the next section have been obtained using $\gamma=220$ and $\epsilon=0.02$ after manual exploration. Similar dynamics were observed for other parameters, but we choose this combination as it resulted in a greater diversity of dynamics for different $\beta$. The seed for the random generation of the correlations has been chosen manually and the procedure for setting them is explained in Appendix~\ref{sec:appendix-corr-initialization}.

Experiments have been performed on several 2 Intel Xeon E5-2683 @ 2.10GHz nodes. The execution of each of the processes for every simulation with the settings commented above has been performed in parallel over dozens of nodes. Each calculation taking approximately 5 minutes per execution with 2 dedicated cores per each process. We dedicated 8GB of memory to each of the processes.


\begin{figure}[ht]
    \begin{tabular}{l c l}
     a)& &b) \\ \includegraphics[width=0.4\linewidth]{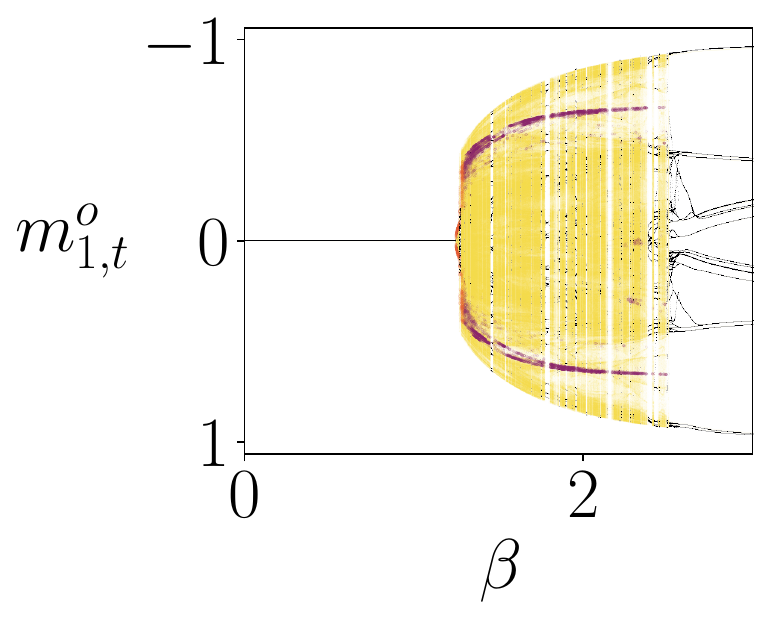} & &\includegraphics[width=0.4\linewidth]{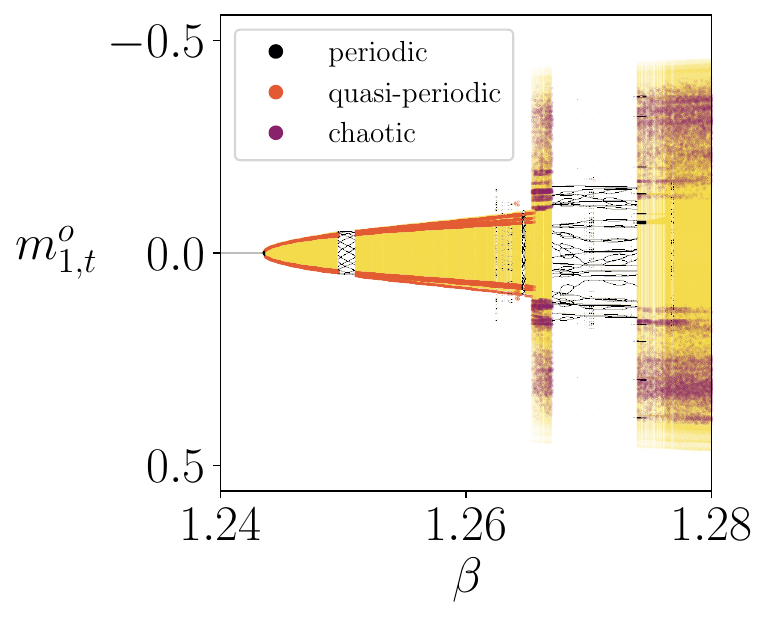}  
    \end{tabular}
    \caption{Bifurcation diagram for $\beta\in[0,3]$ (a) and $\beta \in[1.24,1.28]$ (b). For each $\beta$, we plot all the different points the trajectory has traversed (black points for periodic trajectories, yellow points for the rest).
    For quasi-periodic and chaotic trajectories, points interesecting with the plane $m^o_{2}=0$ are represented in orange and purple, respectively. In both bifurcation diagrams, we have first simulated the highest $\beta$ value and then used the context at the end of the execution as initial condition for other $\beta$. Only the semantic information regarding the mean-field variables is being shown (first term of Eq.~\eqref{eq:mf_o_pe}).}
    \label{fig:bif_diagram}
\end{figure}

\subsection{Nonequilibrium phase transitions}

In this section, we explore the behaviour of the model for different values of the inverse temperature, $\beta=T^{-1}$. In transformers, the next token probabilities depend on a temperature value that modifies the softmax probabilities as in Eq.~\eqref{eq:softmax-output}. We modify temperature values from $0$ to $3$, using $4,001$  values. This is analogous to exploring different temperatures of the softmax output in a standard transformer architecture. 

We generate a bifurcation diagram of the system in two forms. Generally, a bifurcation diagram \cite{strogatz2018nonlinear} depicts all values taken by some variable for a specific parameter (in this case, $\beta$). In the first form of the bifurcation diagram, we simply sample 20,000 points from the steady state of one of the mean-field variables, $m_{1,t}^o$ (other variables behave similarly), and represent the values taken for each $\beta$ as the black and yellow points in Fig.~\ref{fig:bif_diagram}, depending if the attractor is periodic or not. In the second form of the bifurcation diagram, if the attractor is not periodic we plot the points in the trajectory intersecting with the plane $m^o_2=0$ (within a small error value of $0.001$), represented as orange and purple points points in Fig.~\ref{fig:bif_diagram}, depending if the attractor is quasi-periodic or chaotic. Although in future research we will systematically study the attractors, here we discriminate them only by visual inspection and a count of the number of points in each bifurcation diagram.
In Fig.~\ref{fig:bif_diagram}.a and Fig.~\ref{fig:bif_diagram}.b we can see the bifurcation diagrams for $\beta$ values in the ranges $[0,3]$ and $[1.24,1.28]$ respectively. At lower $\beta$, the system falls into periodic attractors, to later develop into a zone with a quasi-periodic behavior that ends up abruptly transitioning into a chaotic regime. We can see that, nonetheless, the different regimes are not stable and small changes in $\beta$ make a large difference in the demonstrated behavior.

\begin{figure}[ht]
\centering
\begin{tabular}{cc}
      \includegraphics[width=.33\linewidth]{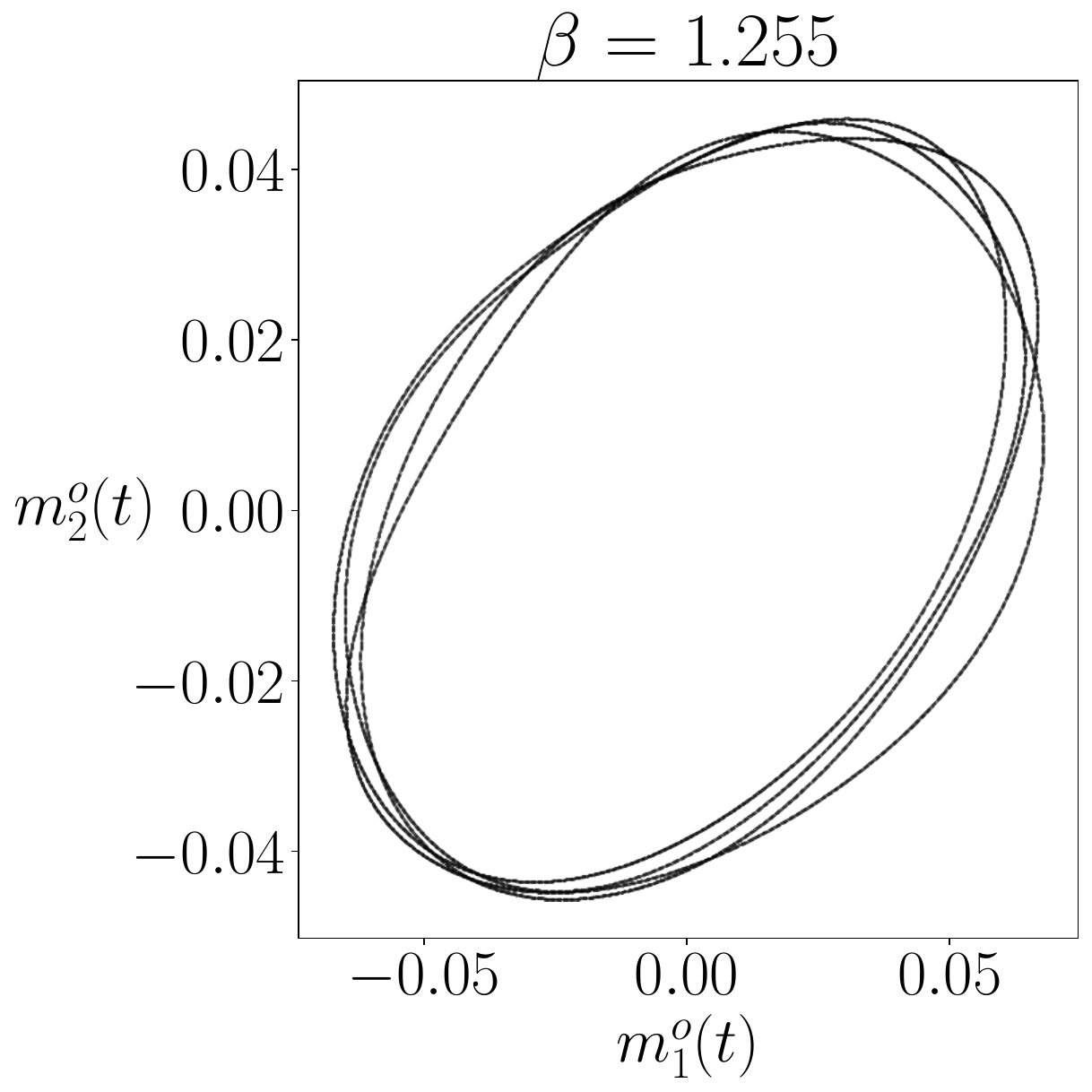}
     &  
     \includegraphics[width=.33\linewidth]{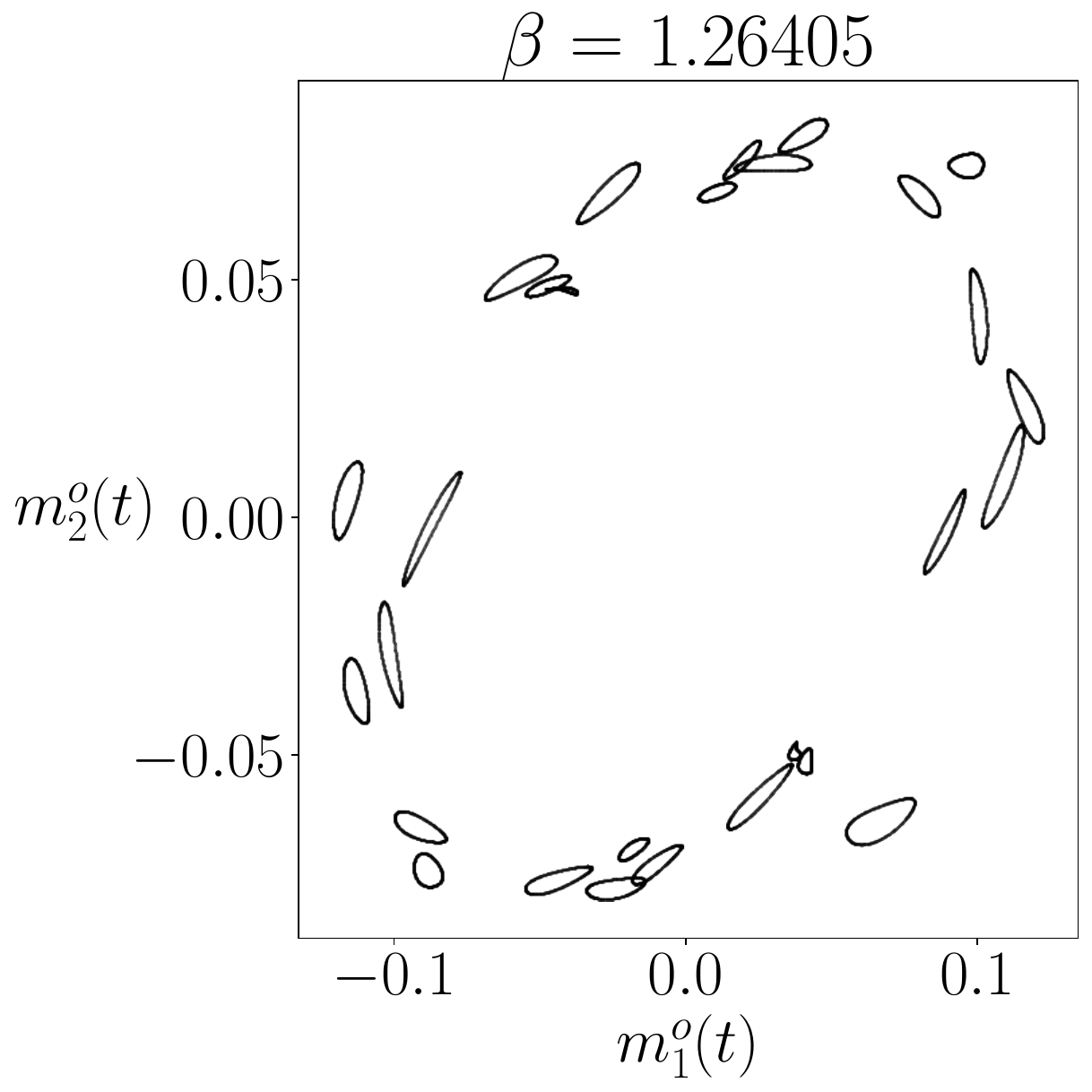}
     \\
     \includegraphics[width=.33\linewidth]{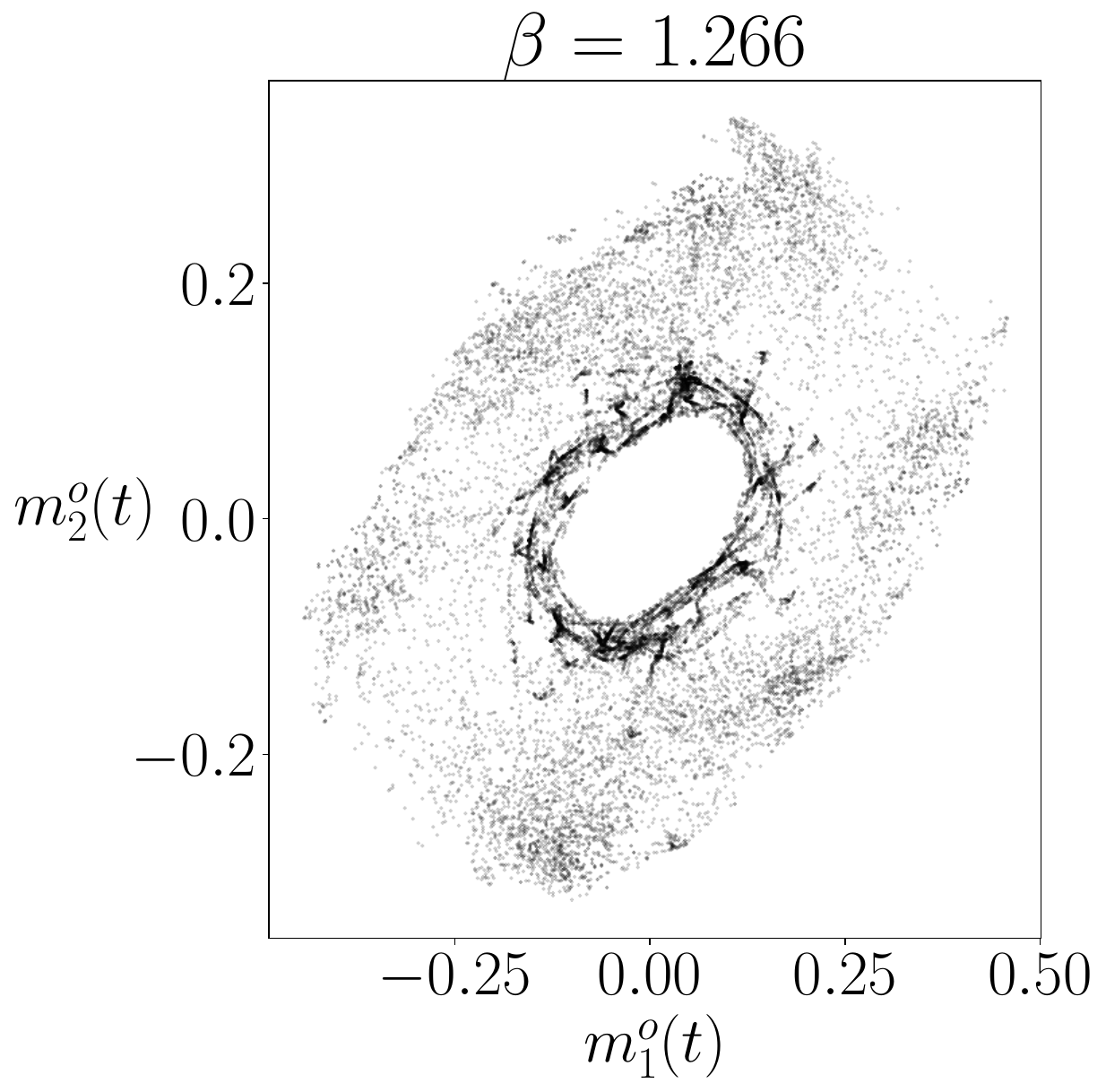}
     & 
     \includegraphics[width=.33\linewidth]{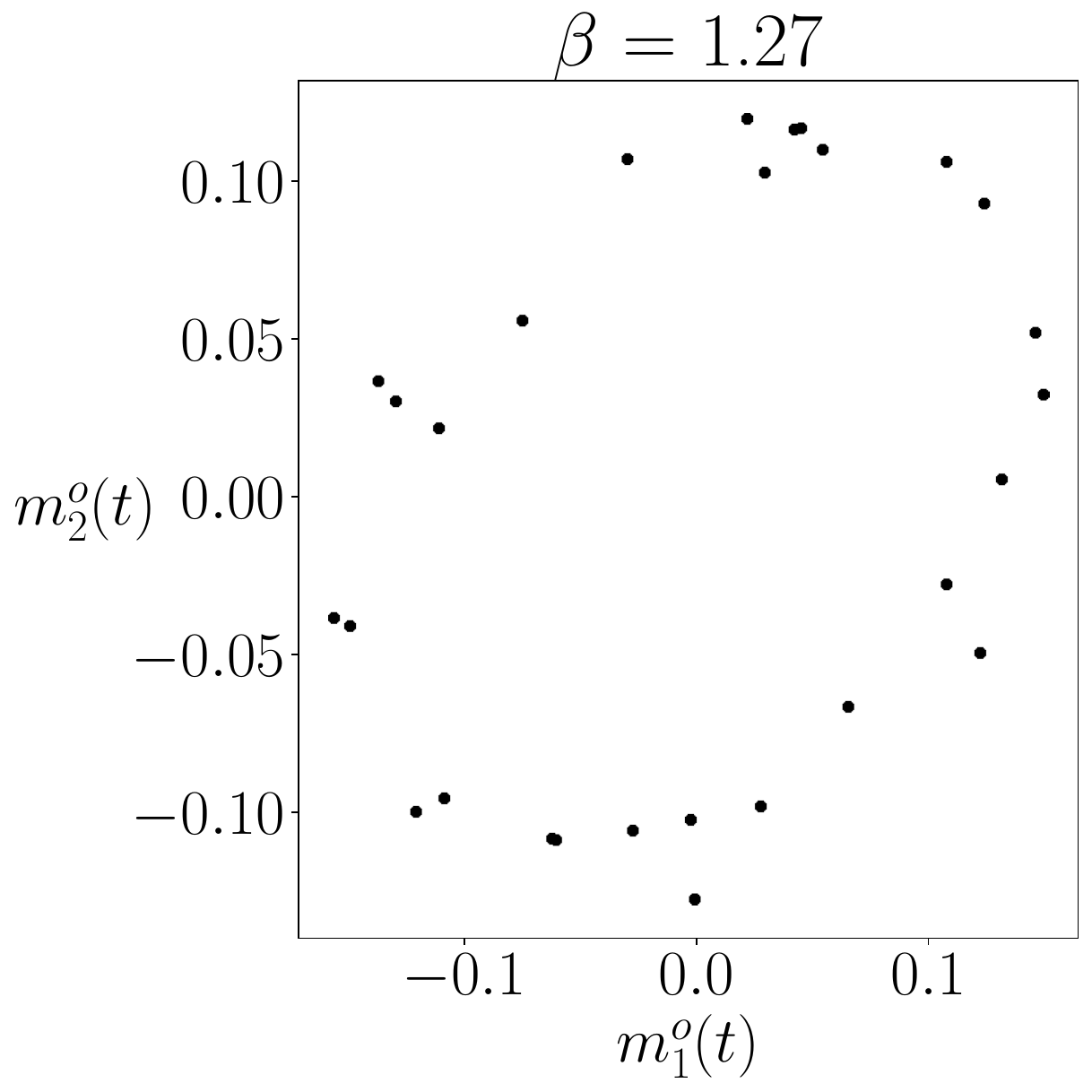}
     \\
     \includegraphics[width=.33\linewidth]{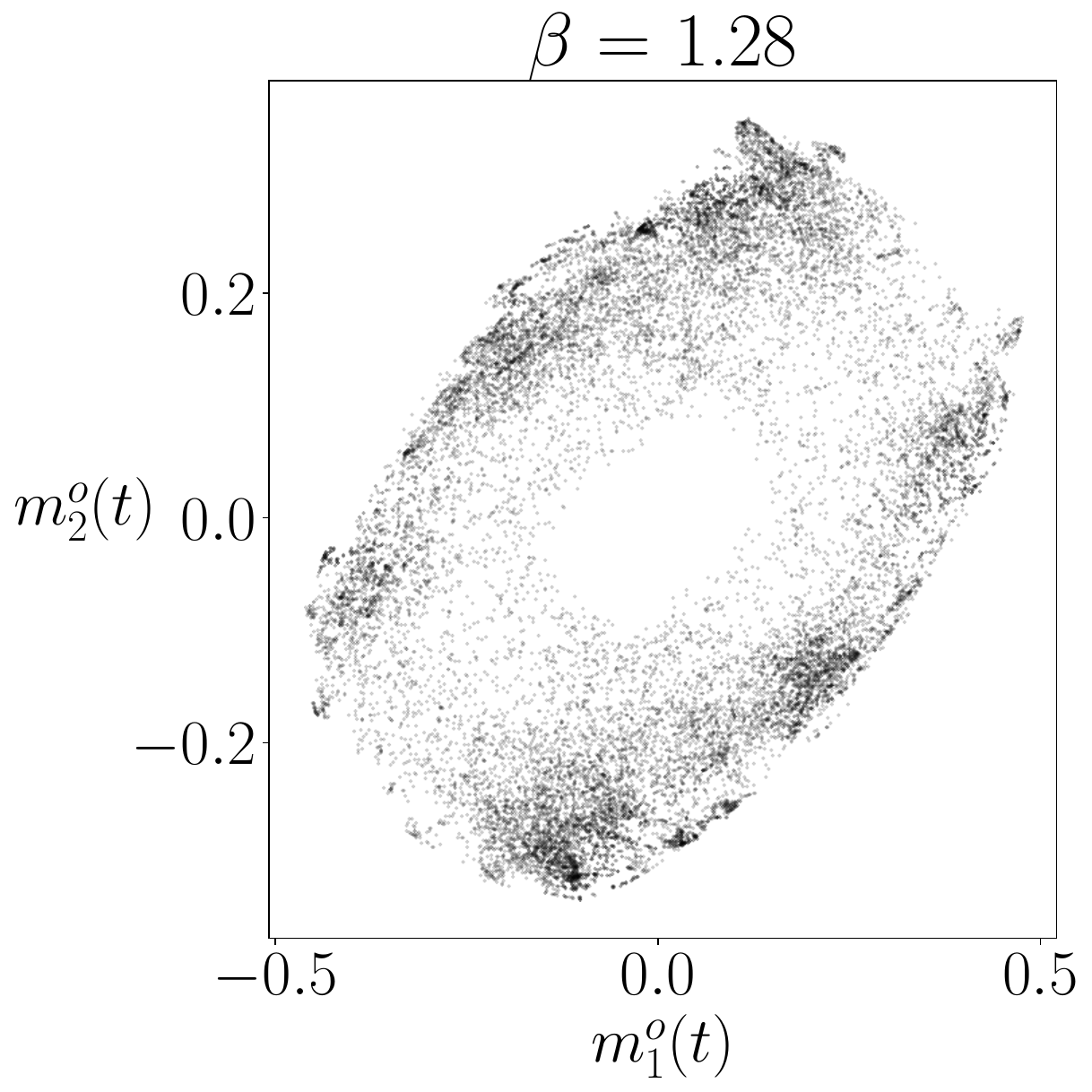}
     &
     \includegraphics[width=.33\linewidth]{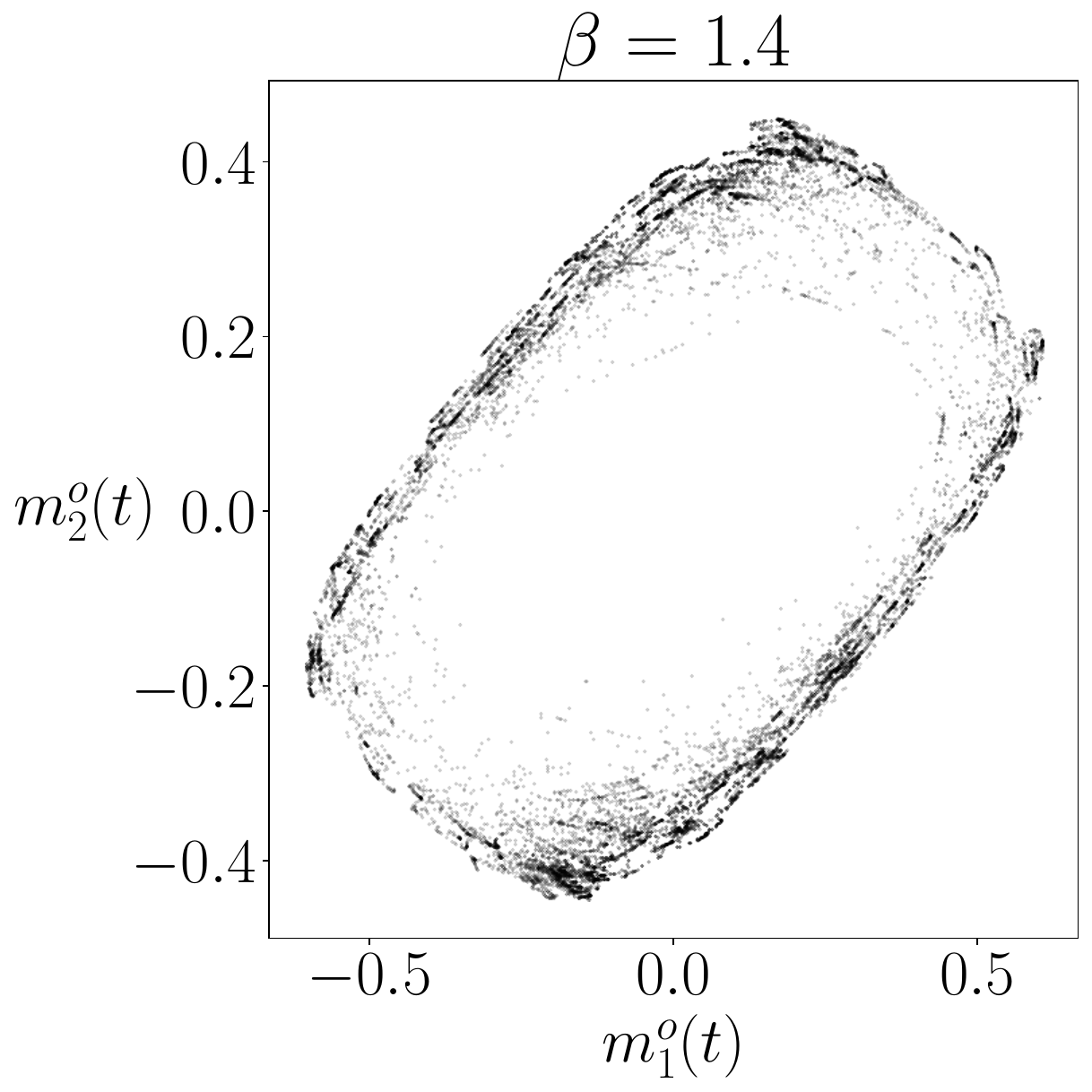}
\end{tabular}
\caption{Trajectories of mean-field variables $m^o_1(t)$ vs. $m^o_2(t)$. Only the semantic part (first term of Eq.~\eqref{eq:mf_o_pe}) is represented for different $\beta$. Points at $\beta=1.27$ are larger to facilitate visualization.}
\label{fig:six-planes}
\end{figure}

In order to visualize this  we have selected values of $\beta$ from the bifurcation diagram and plotted the trajectories of $m^o_1(t)$ against $m^o_2(t)$ in the same plane. Here, we can distinguish between 1) periodic trajectories jumping among a set of fixed points (e.g. $\beta=1.27$ in Fig.~\ref{fig:six-planes}), 2) quasi-periodic cycles jumping between points on a set of continuous curves (e.g. $\beta=1.255$ and $\beta=1.26405$ in Fig.~\ref{fig:six-planes}), or 3) chaotic trajectories (e.g. $\beta=1.266$, $\beta=1.28$ and $\beta=1.4$ in Fig.~\ref{fig:six-planes}).

\subsection{Dynamics and temporal memory}

\begin{figure}[ht]
    \centering
\begin{tabular}{lcl}
         \includegraphics[width=0.4\linewidth]{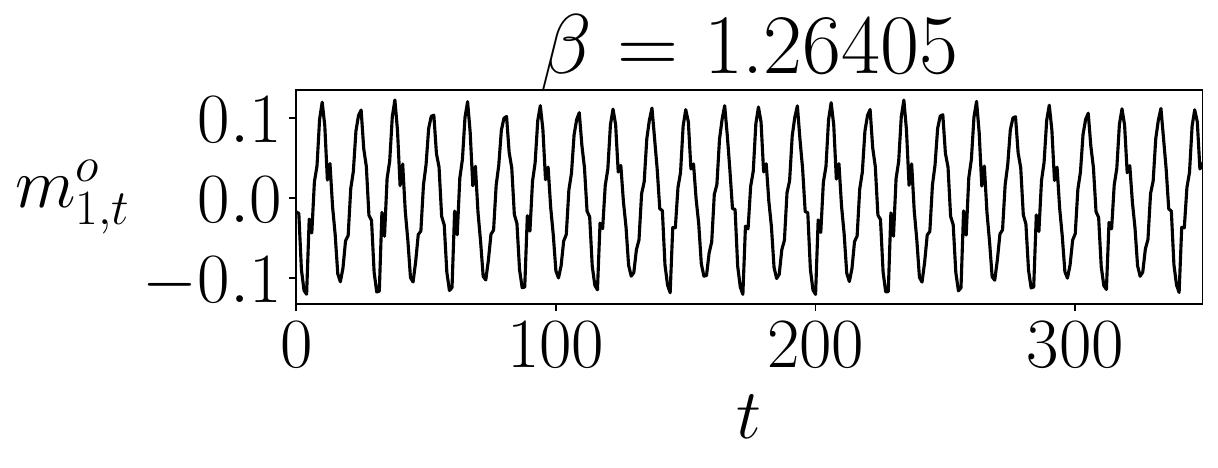}& \hspace{1cm} &\includegraphics[width=0.4\linewidth]{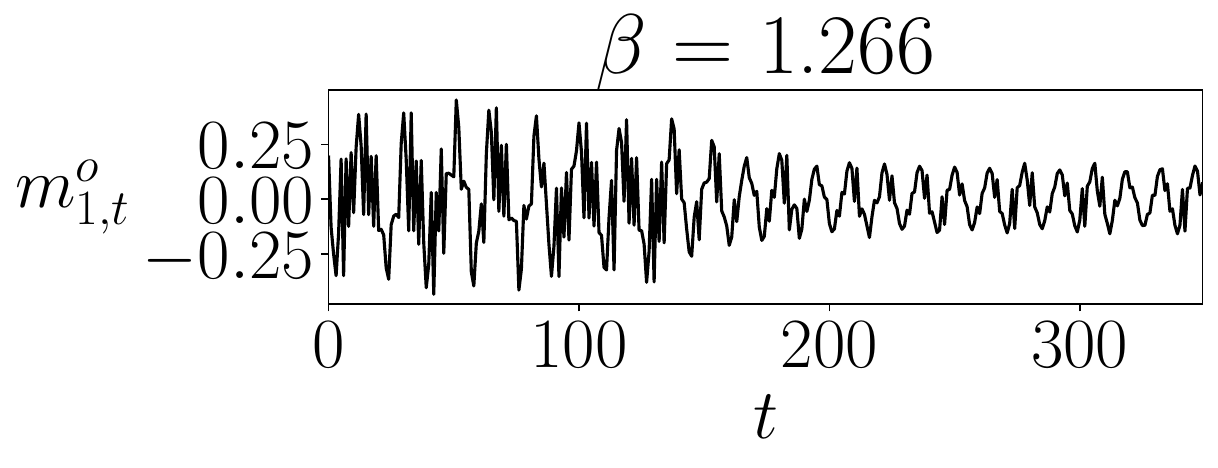}
\\
         \includegraphics[width=0.4\linewidth]{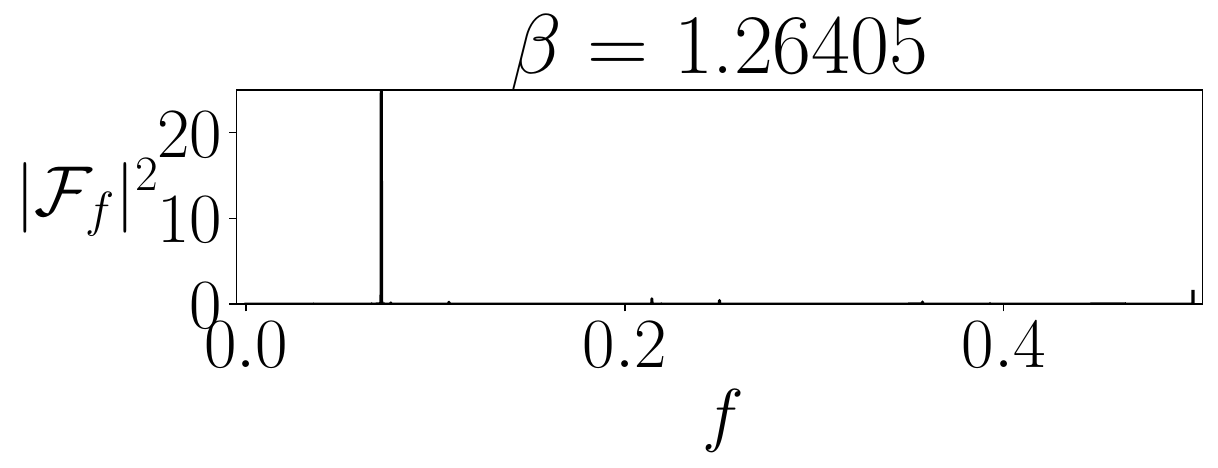}& \hspace{1cm} &\includegraphics[width=0.4\linewidth]{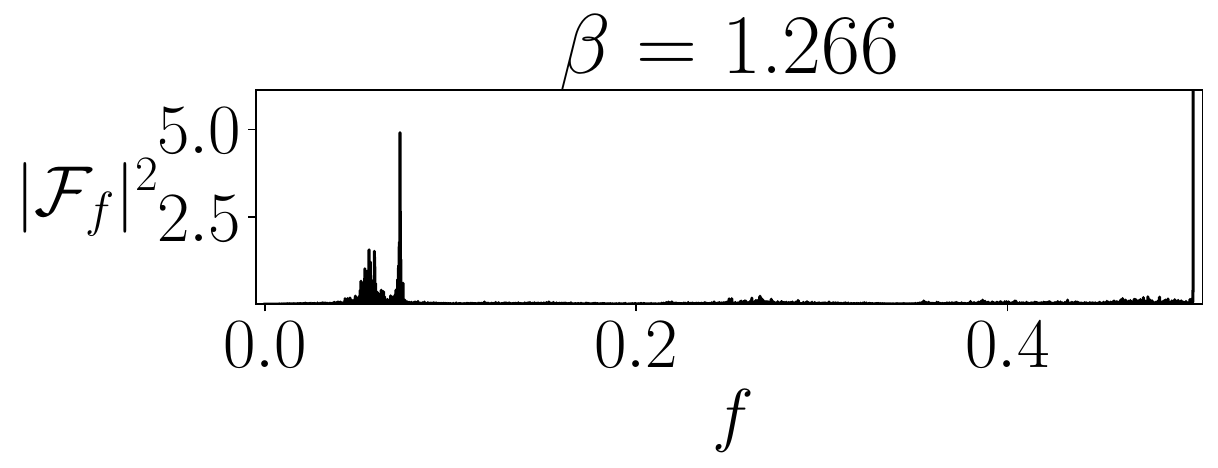}
\\
         \includegraphics[width=0.4\linewidth]{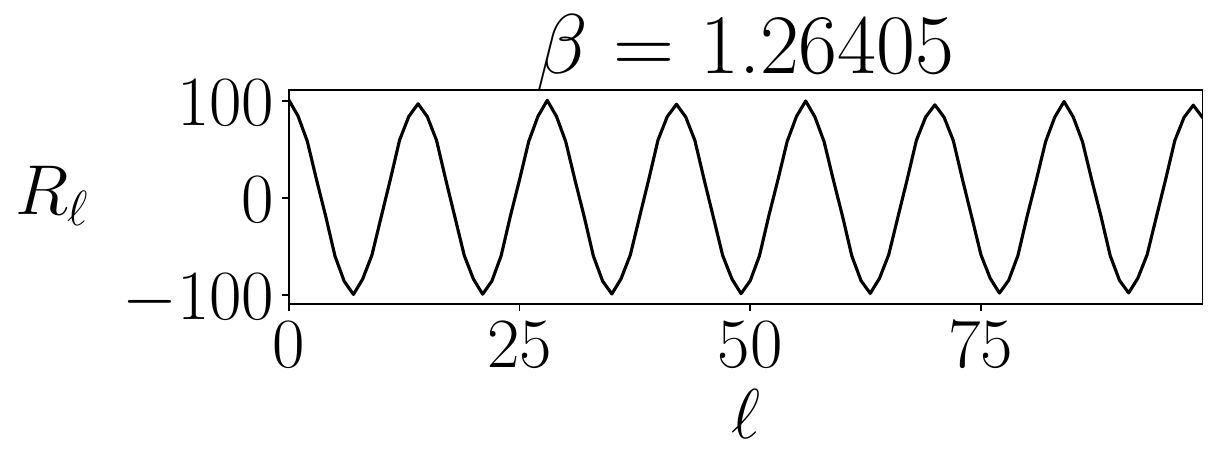} & \hspace{1cm} &\includegraphics[width=0.4\linewidth]{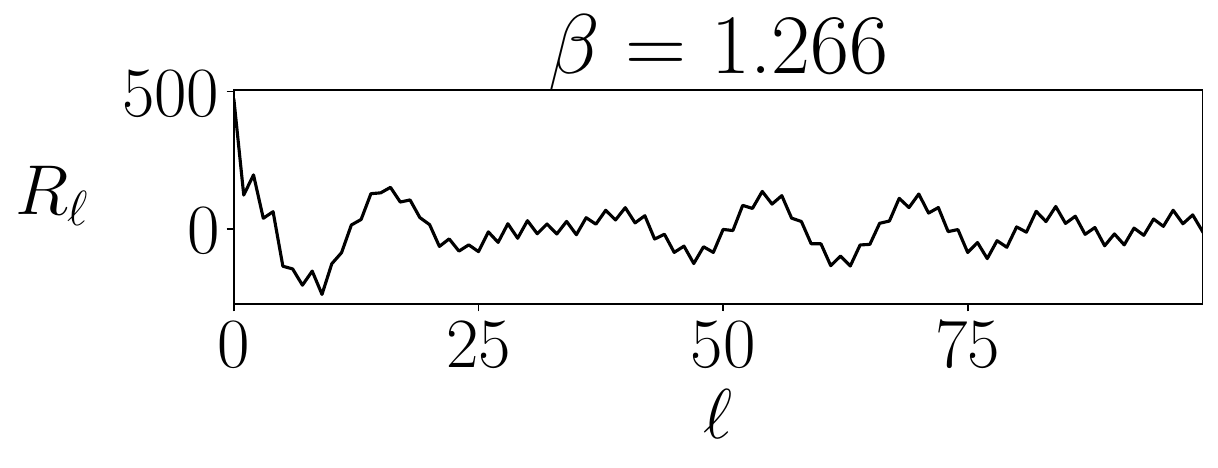}
\end{tabular}
    \caption{Examples of quasi-periodic (left) and chaotic (right) trajectories (top). Mean-field trajectories (first term of first term of Eq.~\eqref{eq:mf_o_pe}) are plotted after 112,800  steps. We also show the discrete Fast Fourier Transform (middle) and autocorrelation function (bottom) of 20,000 samples of $m^o_{1,t}$ at steady-state. Note that the figure at the middle-right has a peak of 15 at $f=0.5$ not shown to better represent other frequencies.}
    \label{fig:chaotic_pseudo_comparison}
\end{figure}

Here, we observe in more detail some of the trajectories in the bifurcation diagram. For both a) periodic and b) chaotic trajectories, we show 1) an example of 1 trajectory, 2) its auto-correlation function $R_\ell(x_{1:T})=\tfrac{1}{T} \sum _{t=0}^{T}x_t x_{t-\ell}$, and 3) its discrete fourier transform  $\mathcal F_f(x_{1:T})=\sum _{t=1}^{T}x_{t} \exp\pr{-\iu 2\pi {\frac{t}{T} f}}$.
In Fig.~\ref{fig:chaotic_pseudo_comparison} (top) we observe an example of a quasi-periodic  and a chaotic trajectory. Observing their Fourier spectra and auto-correlation  function (Figs.~\ref{fig:chaotic_pseudo_comparison}, middle and bottom), we see that all signals have long correlations associated with low-frequency components , specially in the case of the chaotic regime. In the case of the quasi-periodic trajectory, we observe lower frequency components than those associated with the context length ($f=\frac{1}{L}=0.25$).
This sheds light into an often overlooked aspect of attention and transformer models, which is that their memory capacity is not limited to the span of the context window, but the system can display a rich dynamical repertoire inducing low-frequency components increasing significantly the capacity of the model to `remember' previous tokens.

Additionally, quasi-periodic and specially chaotic trajectories --for semantic representations where related tokens are close in the embedding space-- can be interpreted as attractors allowing the system to express a similar structure in different ways. Furthermore, coexistence of different attractors as in Fig.~\ref{fig:chaotic_pseudo_comparison} (top right) shows how a model can switch between different structures driven by chaotic dynamics. Overall, the simple example displayed here shows how transformer-like architectures can easily give rise to rich dynamical structures with nontrivial memory-effects.

\section{Conclusion}

In this article, we present a dynamical mean-field theory for attention networks,  showcasing its application in a simple model combining an attention network with a softmax output token generator. We use techniques from nonequilibrium statistical mechanics to study attention mechanisms as asymmetric Hopfield networks. Our calculations yield an exact result of the statistics over path probabilities in the large limit size. While, for simplicity, we used 1-bit descriptions of weights and tokens, the results can be easily extended to other configurations.

Even for relatively simple configurations, including just three features for each level (keys, queries, values and outputs) and a very limited context (4 tokens), we find a rich dynamical landscape of behaviours, including several phase transitions between periodic, quasiperiodic and chaotic attractors for the mean-field variables. Furthermore, in all cases we observe complicated dynamics with memory effects much beyond what is stored in the context window. Although we did not consider other blocks in the transformer network like feedforward layers and residual connections, we expect that additional nonlinearities and feedback loops will increase the complexity of the observed dynamics.

Even though our examples are limited to a simplified model configuration, our methodology has the potential to provide  insights into the dynamics of transformer models in more realistic setups. Understanding the phase-diagrams expected in general configurations of transformer networks can provide a deeper understanding of the architecture, as well as the characterization of desirable configurations in their parameter space and relevant phase transitions. In addition, our framework can be extended to implement techniques to understand the relation between memory capacity and dynamics for models near saturation \cite{crisanti1988dynamics,coolen2001statistical,aguilera2023nonequilibrium}, i.e. the nonequilibrium equivalent of the celebrated mean-field theory of spin glasses in statistical physics \cite{mezard1984replica}.

Additionally, we hypothesize that our mean-field calculations can provide a cost-effective alternative for computing the statistics of attention layers in transformer models, which could potentially speed-up model training processes by alleviating the computational cost of calculating the gradients of the loss function during learning. Despite the assumption of infinite-large systems in the calculation of mean-field variables, our approach can be refined for more accurate approximations in finite-size networks, for instance using theories from nonequilibrium spin models \cite{mezard2011exact,poc2021inference,aguilera2021unifying}. We expect to explore this idea in future work.

We aspire to contribute in the advancement of methods aimed at enhancing the interpretability of transformer models. The interpretability of features in large language models has raised great interest as well as the identification of circuits involved in key features of the system \cite{templeton2024scaling}.
Through the perspective of nonequilibrium statistical physics, we can elucidate the dynamics of large, heterogeneous networks by describing them through a lower-dimensional set of mean-field variables. Understanding such order parameters results in the identification of key features and patterns in the model's predictions. We speculate that understanding phase transitions in these key features could provide key insights relevant for challenges like alignment problems.

Taken together, these results contribute to the development of an exact analytical theory of the nonequilibrium statistical physics of attention mechanisms and their phase transitions. We envision that contributions in this line can potentially  foster a more systematic integration of theoretical concepts from statistical physics into the domain of generative AI models.

\begin{ack}


We thank Ivan Garashchuk and Iñigo Urteaga for helpful comments on the manuscript.
APL and MA are funded by MA's Junior Leader fellowship from ``la Caixa'' Foundation (ID 100010434,  code LCF/BQ/PI23/11970024). APL and MA acknowledge support by the Basque Government's BERC 2022-2025 program and the Spanish Ministry of Science and Innovation's Severo Ochoa accreditation (CEX2021-001142-S / MICIN/AEI/10.13039/501100011033).
MA is partially supported by John Templeton Foundation (Grant ID 62828), the Basque Government (ELKARTEK 2023 program, project KK-2023/00085) and grant MICIU/AEI /10.13039/501100011033 from the Spanish  Ministry of Science, Innovation and Universities.
\end{ack}







\newpage
{\small
\bibliography{references}
}


\appendix


\newpage

\medskip

\section{Appendix. Full derivation of the generating functional.}\label{sec:appendix-functional}

In this appendix, we develop step by step the functional introduced in Eq.~\eqref{eq:generative-functional}. As a first step, we expand $P(\bm x_{t+1}|\bm A_{t})$ in Eq.~\eqref{eq:annex-mf-f-exp}. To continue, in the last line of Eq.~\eqref{eq:annex-mf-f-delta} we introduce the integral form of the Dirac delta function for each mean-field variable and develop the $\cosh$ from the sum over $\bm  x'$ in the denominator of Eq.~\eqref{eq:annex-mf-f-exp}.
In doing so, we introduce variables
\begin{equation}
 \hat A_{a,t} = \frac{1}{N}\int d\bm m A_{a,t} \prod_{a,t,\alpha} \delta\pr{m_{a,t}^\alpha-\frac{1}{N}\sum_i W_{i,a}^\alpha x_{i,t}}
 \label{eq:annex-hat-A}
\end{equation}
resulting in Eq.~\eqref{eq:mean-field_att}.
In this state, we can finally expand the outer sum over all the values of $\bm x$ into a $\cosh$ and rearrange some terms of the terms in Eq.~\eqref{eq:annex-last-eq-functional}.

\begin{align}
    Z(\bm g) =& \sum_{\bm x_{1:T}} \prod^T_{t=0} P(\bm x_{t+1}|\bm A_{t}) \exp(  \sum_{a,\alpha} g_{a,t}^\alpha \frac{1}{N} \sum_i x_{i,t} W^{\alpha}_{i,a})  \\
    =& \sum_{\bm x_{1:T}} \prod^T_{t=0} \frac{\exp(  \frac{\beta}{N} \sum_a \sum_i x_{i,t+1} W^{o}_{i,a} A_{a,t})}{ \sum_{\bm x_{t'}} \exp( \frac{\beta}{N} \sum_a \sum_i x_{i,t+1} W^{o}_{i,a} A_{a,t})} \exp(  \sum_{a,\alpha} g_{a,t}^\alpha \frac{1}{N} \sum_i x_{i,t} W^{\alpha}_{i,a})  \label{eq:annex-mf-f-exp}\\
    =& (\frac{1}{2\pi})^T \sum_{\bm x_{1:T}} \int d\bm{m} \exp  \Bigg( N\sum^T_{t=0} \sum_a \beta m^o_{a,t+1}  \hat{A}_{a,t} + \sum_\alpha m^\alpha_{a,t} g^\alpha_{a,t} \notag \\ & - \sum_i \log 2 \cosh ( \beta \sum_a W^o_{i,a} \hat{A}_{a,t})  - \sum^T_{t=0} \sum_{a,\alpha} \iu\hat{m}^\alpha_{a,t}( m^\alpha_{a,t} - \frac{1}{N} \sum_i x_{i,t} W^\alpha_{i,a}) \Bigg)\label{eq:annex-mf-f-delta}\\
    =& (\frac{1}{2\pi})^T \int d\bm{m} \exp \Bigg(N\sum^T_{t=0} \sum_a \beta m^o_{a,t+1} \hat{A}_{a,t} + \sum_{\alpha} m^\alpha_{a,t} g^\alpha_{a,t} -\iu\hat{m}^\alpha_{a,t} m^\alpha_{a,t}  \notag   \\
    & + \sum_i \log 2 \cosh (\frac{1}{N}  \sum_{a,\alpha} W^\alpha_{i,a} \iu\hat{m}^\alpha_{a,t} ) - \log 2 \cosh ( \beta\sum_a W^o_{i,a} \hat{A}_{a,t}) \Bigg) \label{eq:annex-last-eq-functional} 
\end{align}

Now, we can compute the solution to this equation by saddle point approximation, a solution which is exact in the large limit size. To this end, we will compute the derivatives of the variables introduced by the Dirac delta to obtain its maximum value. By solving $\frac{\partial Z(\bm g)}{\partial m^\alpha_{a,t}} = 0$ we obtain $\iu\hat{m}^\alpha_{a,t}= \frac{1}{N} g^\alpha_{a,t}$ for $\alpha \neq o$ and

\begin{align}
i \hat{m}^o_{a,t} (\bm g) =& N \beta \hat{A}_{a,t-1} +  g^o_{a,t}.  
\end{align}

We can then substitute the $\hat{m}$ variables  to obtain

\begin{align}
    Z(\bm g) =& \exp \Bigg(\sum^T_{t=0} \sum_i \log 2 \cosh (   \sum_{a} \beta W^o_{i,a} \hat{A}_{a,t-1} + \frac{1}{N} \sum_\alpha W^\alpha_{i,a} g^\alpha_{a,t} ) \notag \\ & - \log 2 \cosh ( \frac{\beta}{N} \sum_a W^o_{i,a} \hat{A}_{a,t}) \Bigg).
\end{align}

Finally, by deriving each of the variables $\bm g$ defined for the functional, we obtain the first order moments of the system:

\begin{align}
        \left.\frac{\partial Z(\bm g)}{\partial g_{a,t}^\alpha}\right|_{\bm g = \bm 0} = m^\alpha_{a,t} =& \frac{1}{N} \sum_i W^\alpha_{i,a}  
             \tanh (  \beta \sum_b  W^o_{i,b}  \hat{A}_{b,t-1}).
\end{align}

\section{Appendix. Mean-field from correlations between weights.}
\label{sec:appendix-correlations}

In this section we introduce a method to simplify the initialization of the the weights in the large limit size. We have shown in Eq.\eqref{eq:mean-field} our mean-field equation for the output. There, we have a sum over an infinite number of elements that we cannot instantiate. However, taking advantage of the $\tanh$ symmetry and by introducing some spins $\sigma$ in a sort of a delta function, we can manipulate this result to obtain a new equation for a small number of patterns.

\begin{align}
m^\alpha_{a,t} =& \frac{1}{N} \sum_i W^\alpha_{i,a} \tanh ( \beta \sum_b  W^o_{i,b}  \hat{A}_{b,t-1} )) \\
=& \frac{1}{N} \sum_i \sum_{\boldsymbol{\sigma}} \prod_b \frac{1}{2}(1 + \sigma_b W^o_{i,b})W^\alpha_{i,a} \tanh ( \beta \sum_b \sigma_b \hat{A}_{b,t-1} )) \\
=& \frac{1}{2^M N} \sum_i \sum_{\boldsymbol{\sigma}} \Big( (1 + \sum_b \sigma_b W^o_{i,b} + \sum_{b<c} \sigma_b \sigma_c W^o_{i,b}  W^o_{i,c} + \sum_{b<c<d} \sigma_b \sigma_c \sigma_d W^o_{i,b}  W^o_{i,c}  W^o_{i,d} + \cdots)  \notag \\ 
& \cdot W^\alpha_{i,b} \tanh ( \beta  \sum_b   \sigma_b \hat{A}_{b,t-1} ) \Big) \label{eq:bef_simp}\\
=& \frac{1}{2^M N} \sum_i \sum_{\boldsymbol{\sigma}} \Big( (\sum_b \sigma_b W^o_{i,b} W^\alpha_{i,a} + \sum_{b<c<d} \sigma_b \sigma_c \sigma_d W^o_{i,b}  W^o_{i,c}  W^o_{i,d} W^\alpha_{i,a}+ \cdots) \notag \\
&  
\cdot \tanh ( \beta \sum_b   \sigma_b \hat{A}_{b,t-1} ) \Big) \label{eq:after_simp} \\ 
=& \frac{1}{2^M} \sum_{\boldsymbol{\sigma}} \Big( (\sum_b \sigma_b \ang{W^o_b W^\alpha_a}_i + \sum_{b<c<d} \sigma_b \sigma_c \sigma_d \ang{W^o_{i,b} W^o_{i,c} W^o_{i,d} W^\alpha_{i,a}}_i   + \cdots) \notag \\ &
\cdot \tanh ( \beta \sum_b \sigma_b \hat{A}_{b,t-1} )\Big) .\label{eq:mean_field_corrs}
\end{align}

Thanks to this transformation, we no longer have a sum defined over an infinite number of components $W^\alpha_{i,a}$, but we can just initialize the correlations between the different sets of patterns $\ang{W^o_b,W^\alpha_a}_i$ and $\ang{W^o_{b},W^o_{c},W^o_{d},W^\alpha_a}_i$ .

\section{Appendix. Initialization of correlations between weights.}
\label{sec:appendix-corr-initialization}

Since our purpose in this paper is to explore what basic self-attention networks setups can achieve, we define very simple patterns for the correlations between weights described in Appendix~\ref{sec:appendix-correlations}. To this end, we define our matrices $W^\alpha_{i,a}$ where each element $i$ belongs to one of three equally long segments. Then, we set $W^\alpha_{i,a}\in\{\pm1\}$ with a binomial distribution with probability $p=0.5$, which leads the correlations to be set to $\ang{W^o_b,W^\alpha_a}_i\in \{\pm 1, \pm \frac{1}{3}\}$ and $\ang{W^o_{b},W^o_{c},W^o_{d},W^\alpha_a}_i \in \{\pm 1, \pm \frac{1}{3}\}$.

\end{document}